\RequirePackage{ifpdf}
\documentclass[letterpaper]{JHEP3}
\usepackage{amsmath}
\usepackage{epsfig}
\usepackage{subfigure}

\newcommand{\roughly}[1]{\mathrel{\raise.3ex\hbox{$#1$\kern-0.85em
\lower1ex\hbox{$\sim$}}}}

\def\pd{\partial}

\def\endignore{}
\def\ignore #1\endignore{} 

\def\cO{{\cal O}}

\def\cR{{\cal R}}

\def\BLF{{\scriptscriptstyle BLF}}

\newbox\charbox
\newbox\slabox
\def\slsh#1{{      
        \setbox\charbox=\hbox{$#1$}
        \setbox\slabox=\hbox{$/$}
        \dimen\charbox=\ht\slabox
        \advance\dimen\charbox by -\dp\slabox
        \advance\dimen\charbox by -\ht\charbox
        \advance\dimen\charbox by \dp\charbox
        \divide\dimen\charbox by 2
        \raise-\dimen\charbox\hbox to \wd\charbox{\hss/\hss}
        \llap{$#1$}
}}

\def\exd{{\hbox{d}}}
\def\d{\exd}

\def\bea{\begin{eqnarray}}
\def\eea{\end{eqnarray}}
\def\be{\begin{equation}}
\def\ee{\end{equation}}

\def\ssB{{\scriptscriptstyle B}}

\def\ssM{{\scriptscriptstyle M}}
\def\ssN{{\scriptscriptstyle N}}
\def\ssP{{\scriptscriptstyle P}}

\def\nn{\nonumber}

\def\d{\mathrm{d}}

\def\({\left(}
\def\){\right)}

\def\pref#1{(\ref{#1})}

\def\d{\mathrm{d}}

\numberwithin{equation}{section}

\title{A Problem With $\delta$-functions:\\ Stress-Energy Constraints on Bulk-Brane Matching \\ (with comments on {\tt arXiv:1508.01124})}

\author{C.P.~Burgess,$^{1,2}$ Ross~Diener${}^{1,2}$ and M. Williams$^{3}$ \\
$^1$ Physics \& Astronomy, McMaster University,
Hamilton, ON, Canada, L8S 4M1\\
$^2$ Perimeter Institute for Theoretical Physics, Waterloo, ON, Canada N2L 2Y5\\
$^3$ Instituut voor Theoretische Fysica, KU Leuven,
B-3001 Leuven, Belgium
}

\date{\today}

\abstract { We critically assess a recent assertion \cite{DilatonDeltas} concerning using $\delta$-functions to analyze how higher-codimension branes back-react on their environment. We also briefly summarize the state of the art: describing how stress-energy balance dictates the components of off-brane stress energy in terms brane tension; how this can modify the standard tension/defect-angle relation for codimension-two sources when dilatons are present; and how it all relates to extra-dimensional searches for a small cosmological constant. }

\begin{document}

\section{Background}

In $D$ spacetime dimensions the trace-reversed Einstein equations read
\be
 R_{\ssM \ssN} = \kappa^2 \left( T_{\ssM \ssN} - \frac{1}{d} \, {T^\ssP}_\ssP \, g_{\ssM \ssN} \right) \,,
\ee
with $d = D-2$. This suggests that the $d$-dimensional curvature, $R_{\mu\nu}$, for a maximally symmetric $d$-dimensional source (with stress energy $T_{\mu\nu} = \tau \, g_{\mu\nu}$) can be independent of $\tau$ since
\be \label{4DR}
 R_{\mu\nu} = - \frac{\kappa^2}{d}\, {T^m}_m \, g_{\mu\nu} \,,
\ee
where we split the $D$ directions into $d$- and 2-dimensional subsets: $\{ x^\ssM \} = \{ x^\mu, x^m \}$. This is ultimately why (for instance) relativistic strings can be flat for any value of their tension when embedded into asymptotically flat 4D spacetimes \cite{Vilenkin}.

Several brane-world approaches to the cosmological constant problem have tried to build upon this observation \cite{CLP, CG, SLED}, using $(D,d) = (6,4)$ to explain why cosmology sees such small curvatures, $R_{\mu\nu}$, despite the expectation that the known elementary particles should produce a large 4D vacuum energy, $\tau$. Besides allowing 4D energy to curve unobservable higher dimensions (rather than those seen by cosmologists) higher dimensions are also useful because in them supersymmetry can also forbid a higher-dimensional cosmological constant, leading to the Supersymmetric Large Extra Dimensions (SLED) proposal\footnote{For a recent review see \cite{LesHouches}.} \cite{SLED}. As eq.~\pref{4DR} shows, the proposal hinges on properly identifying the {\em off-brane} stress-energy, $T_{mn}$, for both bulk fields and brane-localized sources.

Although this is a tempting line of argument, people remain (rightly) skeptical \cite{HPT,GP,VC}, pointing out many things that could generically go wrong, often  boiling down to variants of a generic `no-go' argument \cite{Wbgnogo} that identifies scale-invariance as usually playing an important role, and shows why this leads to a problem. What is more difficult is to pin this down precisely, to identify whether or not the generic arguments contain loopholes and to determine systematically on which parameters low-energy curvatures depend.\footnote{For instance, although \cite{GP} identifies a problem if one assumes back-reaction comes only from the defect angle induced by a brane, it is in the end inconclusive because it ignores equally large contributions branes induce for other features of the extra-dimensional geometry \cite{BLFFluxQ, Companion, Companion2} that must also be included to reliably infer how the system really behaves.}

We argue here why ref.~\cite{DilatonDeltas} is similarly inconclusive, using its appearance to highlight how a commonly used $\delta$-function technique for coupling branes to the bulk can be misleading (or at best insufficiently precise to resolve the issues involved) and how to do better. In 6D supergravity the curvature (and so effective cosmological constant) on a space-filling brane turns out to be directly linked to the near-brane radial derivative \cite{ScaleLzero, 6Dnogonot}, $\phi'$, of one of the bulk fields (the dilaton). A central question therefore asks how $\phi'$ is related to the properties of the source branes and under what circumstances can it and the on-brane curvature be small. A detailed answer to this is given in \cite{Companion2}, but ref.~\cite{DilatonDeltas} claims to be able to do so using a much simpler $\delta$-function technique, which sidesteps the ostensibly superfluous complications of \cite{Companion2}.
We repeat --- then critique --- this argument to underline a trap into which one can easily fall.

\section{A critique}
\label{sec:deltafail}

The basic problem is to determine how a specific source brane, described say by an action\footnote{Much of the most interesting discussion in \cite{Companion, Companion2} is about localized flux carried by the branes, but this is not important for the $\delta$-function ambiguity described here so we do not introduce this complication.}
\be \label{Sloc}
 S_{\rm loc} = -  \int \d^4 \xi \sqrt{-\gamma} \, T(\phi)  \,,
\ee
couples to the bulk fields, described say by
\be
 S_{\ssB} = - \int \d^6 x \sqrt{-g} \left[ \frac{1}{2 \kappa^2} \left[ \cR + \left( \partial \phi\right)^2 \right]  + \frac14 \, e^{-\phi} A_{\ssM \ssN}^2 + \frac{2 g^2}{\kappa^4}\, e^{\phi} \right] \,,
\ee
with $A_{\ssM \ssN}$ representing a Maxwell field strength and $\gamma_{\mu\nu} = g_{\ssM\ssN} \partial_\mu z^\ssM \partial_\nu z^\ssN$ denoting the induced metric on the brane, whose position is $x^\ssM = z^\ssM(\xi)$ (see \cite{Companion2} for notational conventions). An important role in this system is played by the invariance of the classical bulk equations under the rigid rescalings $g_{\ssM \ssN} \to c\, g_{\ssM\ssN}$ and $e^{-\phi} \to c\, e^{-\phi}$.

The problem when coupling $S_{\rm loc}$ to $S_\ssB$ is that the source action is lower-dimensional than the bulk action, and this difference must be bridged to infer the effects of the source on the bulk. There are two related ways to proceed. What we call the `$\delta$-function' procedure simply promotes the lower-dimensional source action to a 6D action by introducing a localization function
\be \label{eq:deltaintro}
 \widehat S_{\rm loc} = - \int d^{6} x \sqrt{-g} \,  T(\phi) \, \left[ \frac{ \delta(x - z(x)) }{\sqrt{h}} \right] \,,
\ee
which specializes to a metric $\exd s^2 = g_{\mu\nu}(x,y)\,\exd x^\mu\exd x^\nu + h_{mn}(y) \,\exd y^m \exd y^n$ and the $\delta$-function is a scalar density that, for any $F$, satisfies
\be
 \int \d^2 y \, \delta(y-z) \, F(x,y) = F(x,z) \,,
\ee
(without metrics) as usual. The field equations are then computed by adding eq.~\pref{eq:deltaintro} to $S_\ssB$ and varying the bulk fields in the usual way. In particular, the crucial stress-energy component, $T_{mn}$, is obtained by differentiating $\widehat S_{\rm loc}$ with respect to $g_{mn}$.

The subtle part is deciding how $\delta(y-z)$ depends on the fields, and it is tempting to assume it does not depend on them at all (as \cite{DilatonDeltas} effectively does, as we see below). Although plausible at first glance, this assumption is suspicious for the extra-dimensional metric given that $\delta(y-z)$ is designed to discriminate points according to their proper distance from the point $y=z$. We show here that $\delta$ really must depend on $g_{mn}$, and --- more importantly --- how this dependence can be simply derived in terms of $T(\phi)$ from the stress-energy balance of the UV physics that the $\delta$-function is meant to represent.

To see how, we use the metric ansatz $\exd s^2 = W^2(\rho) \,\check g_{\mu\nu}(x)\,\exd x^\mu\exd x^\nu + \exd \rho^2 + B(\rho)^2\,\exd\theta^2$ (with $\check g_{\mu\nu}(x)$ a maximally-symmetric 4-metric) and find the Maxwell equation integrates to give the nonzero component $A_{\rho\theta} = {Q\,B\,e^\phi}/{W^4}$ with integration constant $Q$. Assuming $\delta(y-z)$ to be dilaton independent, the dilaton field equation reads
\be \label{eq:dilaton}
 \Box \phi = \frac{1}{BW^4} \Bigl( BW^4 \phi' \Bigr)' = \kappa^2 e^\phi \left( \frac{2 g^2}{\kappa^4} - \frac{Q^2}{2W^8} \right) + \sum_b \kappa^2 T^\prime_b(\phi) \left[ \frac{\delta(y-z_b)}{B} \right] \,,
\ee
where primes denote derivatives with respect to the appropriate arguments --- {\em ie} $\rho$ for $B$, $W$ and $\phi$ and $\phi$ for $T_b$ --- and the sum is over any source branes present. Assuming $\delta(y-z)$ is metric independent gives the Einstein equations
\be \label{eq:4dEins}
 - \frac{1}{\kappa^2} \, \frac{\left[ B^\prime W^4 \right]^\prime }{B W^4} =  e^\phi \left( \frac{3 Q^2 }{4 W^8 } + \frac{g^2 }{\kappa^4}\right)  + \sum_b T_b(\phi) \left[ \frac{ \delta(y - z_b)}{B } \right] \,,
\ee
\be \label{eq:bulkdiffEins}
\frac{\check R}{ W^2} + \frac{\left[ B (W^4)^\prime \right]^\prime}{ B W^4 } = 2 \kappa^2 e^{\phi} \left( \frac{Q^2}{ 2 W^8} - \frac{2 g^2}{\kappa^4} \right)
\ee
together with the `constraint'
\be \label{eq:const}
 8 \left( \frac{B' W'}{BW} \right) + \frac{\check R}{ W^2} +  12 \left( \frac{W'}{W} \right)^2  - \left( \phi^\prime \right)^2  =  2 \kappa^2  e^{\phi} \left( \frac{Q^2}{2 W^8} - \frac{2 g^2}{\kappa^4}  \right) \,.
\ee
These agree with those found in \cite{DilatonDeltas}.

\subsection*{Boundary conditions}

The connection between curvature and $\phi'$ comes from summing \pref{eq:bulkdiffEins} with twice \pref{eq:dilaton}, multiplying by $BW^4$ and integrating the result over the transverse directions, giving
\be \label{Rphi'}
 \check R \int \exd^2y \, BW^2 + \int \exd^2 y \Bigl[ 2 BW^4 \phi' + (B W^4)' \Bigr]'  = \sum_b 2\kappa^2 \int \exd^2y \; W^4  T^\prime_b(\phi) \; \delta(y-z_b) \,.
\ee
On one hand, integrating this over a region completely exterior to the branes, ending an infinitesimal distance, $\rho = \epsilon$, away, shows that $\check R$ vanishes if $BW^4(\phi' + 2 W'/W)_{\rho = \epsilon}$ vanishes near both branes. On the other hand $\phi'$, $W'$ and $B'$ at $\rho = \epsilon$ can be found by integrating the above field equations over a complementary infinitesimal region, $(\rho - \rho_b) \le \epsilon$, that just barely includes each brane, and then taking the limit $\epsilon \to 0$. This leads to the other way to connect the brane and bulk actions: relating the near-brane boundary conditions for bulk fields directly to the derivative of the 4D action, \pref{Sloc}.

For example, performing this operation on the dilaton field equation gives
\be \label{eq:dilatonBC}
 2 \pi  \lim_{\epsilon \to 0 } \Bigl( BW^4 \phi' \Bigr)_{\rho = \rho_b+\epsilon} = \lim_{\epsilon \to 0 } \int_{\rho-\rho_b < \epsilon} \d^2 y \sqrt{- g }\,  \Box \phi = \kappa^2 W_b^4 T^\prime_b(\phi_b) \,,
\ee
where $\phi_b = \phi(\rho_b)$ and similarly for $W_b$. Making sense of this equation requires knowing how each side behaves as $\rho \to \rho_b$, but this is determined by the bulk field equations which in the near-brane limit (see Appendix) give power-law solutions \cite{6DdS}
\be \label{Kas}
 e^\phi \propto \hat\rho^{z_b} \,, \quad B \propto \hat\rho^{\beta_b} \quad \hbox{and} \quad W \propto \hat\rho^{w_b} \,,
\ee
where $\hat \rho = \rho - \rho_b$ and the powers $\beta_b$, $w_b$ and $z_b$ must satisfy the `Kasner' conditions
\be \label{Kasnerc}
 4w_b + \beta_b =
 4 w_b^2 + \beta_b^2 + z_b^2 = 1 \,.
\ee
This leaves one independent combination, and it is this that is fixed by the boundary conditions in terms of $T(\phi)$.

Notice that in general (unless $z_b = 0$) $\phi$ diverges logarithmically as $\rho \to \rho_b$, and \pref{Kas} shows how this can also lead to singular curvature in this limit. Notice however that the condition $\beta_b + 4w_b = 1$ implies that the left-hand side of \pref{eq:dilatonBC} in all cases has a finite limit as $\epsilon \to 0$. Naively the same need not also be true of the right-hand side, depending on the functional form of $T(\phi)$. But as shown in \cite{Companion2} (see also \cite{HiggsRen}) consistency is always restored by the $\epsilon$-dependence implied by the renormalization of the brane-bulk couplings \cite{ClassRenorm} required even at the classical level to ensure physical properties remain finite as $\epsilon \to 0$. In particular these relations ensure that $e^\phi$ and $W$ are smooth enough at the brane position that the integral over the potential and Maxwell contributions to the dilaton field equation generally do not survive the limit $\epsilon \to 0$.

To the extent that one trusts eqs.~\pref{eq:4dEins} the boundary condition for the metric function $B$ follows similarly
\be \label{eq:BBC}
 2 \pi \lim_{\rho \to \rho_b} \left[ W^4 B^\prime - 1  \right] = - \kappa^2 W_b^4 T_b(\phi_b)  \qquad \hbox{(tentative)} \,,
\ee
(which is the usual relation between tension and defect angle.) Eq.~\pref{eq:bulkdiffEins} similarly gives a trivial boundary condition for the warp factor
\be \label{eq:WBC}
 2 \pi \lim_{\rho \to \rho_b}  [ B (W^4)^\prime ]= 0 \qquad \hbox{(tentative)} \,.
\ee

\subsection*{$\delta$-function failure}

We now can see more precisely what is wrong with taking the $\delta$-function independent of bulk fields. As argued in \cite{UVCaps} the problem is that the boundary conditions derived generically do not satisfy the constraint equation, \pref{eq:const}. To see this multiply \pref{eq:const} through by $(W^4 B)^2$ and evaluate the result in the limit $\rho \to \rho_b$. This gives
\be \label{newconstr}
 2 \lim_{\rho \to \rho_b}\left[ W^4 B^\prime  \right] \left[ B (W^4)^\prime \right] + \frac{3}{4} \lim_{\rho \to \rho_b} \left[ B (W^4)^\prime \right]^2 = \lim_{\rho \to \rho_b} \left[W^4 B \phi^\prime \right]^2 \,,
\ee
which is a nontrivial relation between the near-brane boundary conditions for the fields $B$, $W$ and $\phi$. Furthermore, each factor is finite as $\rho \to \rho_b$ and so \pref{eq:WBC} combined with \pref{newconstr} implies
\be
 \lim_{\rho \to \rho_b} \left[W^4 B \phi^\prime \right] = 0 \,.
\ee
This result is in general inconsistent with the dilaton boundary condition \pref{eq:dilatonBC}, which indicates (supported by the numerics of \cite{Companion2}, which constructs explicit UV completions for the brane) that choices for $T(\phi)$ should exist that are consistent with nonzero $\lim_{\rho \to \rho_b} B W^4 \phi'$.

What is going on? The problem is the assumption of metric-independent $\delta(y)$. The constraint equation breaks down ultimately because this assumption is inconsistent with stress-energy balance, which is satisfied for {\em any} UV completion of the brane. A similar breakdown is also seen if the brane is regularized by giving it substructure (such as representing it as a codimension-one ring \cite{TP, UVCaps}) if care is not taken to stabilize the ring's radius since the failure of the radial stresses to balance implies an inconsistency with the radial Einstein equation.

\subsection*{Determining the field-dependence of $\delta(y)$}

If $\delta(y)$ must depend on $g_{mn}$, how is this dependence determined?
We here recap the arguments \cite{Companion2, UVCaps} that show how this can be inferred using the constraint \pref{newconstr}. To see how, we leave the derivative
\be
 \frac{\pd \,\delta(y)}{\pd g_{mn}} = C^{mn}\,\delta(y) \,,
\ee
unspecified in the field equations, which for rotationally invariant sources gives two independent components:\footnote{The counting is the same for rotationally invariant sources with higher codimension, so the arguments given here suffice to determine $C_{mn}$ in this case as well.} ${C^\rho}_\rho$ and ${C^\theta}_\theta$. This modifies eqs.~\pref{eq:4dEins}--\pref{eq:const} to become
\be \nn
 \frac{\left[ B^\prime W^4 \right]^\prime }{B W^4} = - \kappa^2 \left[ e^{\phi}\left( \frac{3 Q^2 }{4 W^8 } + \frac{g^2 }{\kappa^4} \right) + \sum_b \frac{T_b(\phi)}{B} \delta(y-z_b) \left( 1 +\frac{3}{2} C^{\theta}{}_{\theta} - \frac{1}{2} C^{\rho}{}_{\rho}  \right)  \right] \\ \nn
\ee
\be
 \frac{\check R}{ W^2} + \frac{\left[ B (W^4)^\prime \right]^\prime}{ B W^4 } = 2 \kappa^2 \left[e^{\phi} \left( \frac{Q^2}{ 2 W^8} - \frac{2 g^2}{\kappa^4} \right) - \sum_b\frac{T_b(\phi)}{B}\delta(y-z_b) \left( C^{\theta}{}_{\theta} + C^{\rho}{}_{\rho} \right) \right]
\ee
\be
 8 \left( \frac{B' W'}{BW} \right) + \frac{\check R}{ W^2} +  12 \left( \frac{W'}{W} \right)^2 - \left( \phi^\prime \right)^2  =  2 \kappa^2  \left[ e^{\phi} \left( \frac{Q^2}{2 W^8}  - \frac{2 g^2}{\kappa^4} \right) + 2 \sum_b\frac{T_b(\phi)}{B}\delta(y-z_b) C^{\rho}{}_{\rho}   \right] \,,\nn
\ee
which we integrate as before to relate near-brane asymptotics to brane properties.

Integrating the third of these equations over the disc $|\rho-\rho_b|<\epsilon$ and taking the limit $\epsilon \to 0$ gives
\be
 C^{\rho}{}_{\rho} = 0 \,,
\ee
because the equation is smooth in the near-source limit.\footnote{The asymptotic form \pref{Kas} actually gives terms that diverge but these cancel due to \pref{Kasnerc}.} Integrating the other two similarly modifies the tension/defect-angle boundary conditions eqs.~\pref{eq:BBC} into
\be \label{eq:BBCnew}
 2 \pi \lim_{\rho \to \rho_b} \left[ W^4 B^\prime - 1  \right] = - \kappa^2  W_b^4T_b(\phi_b) \left(1+ \frac{3C}{2} \right) \,,
\ee
while \pref{eq:WBC} becomes
\be \label{eq:WBCnew}
 2 \pi \lim_{\rho \to \rho_b}  [ B (W^4)^\prime ]= - 2\,\kappa^2 W_b^4 T_b(\phi_b) \,C\,,
\ee
where $C = C^{\theta}{}_{\theta}$. The constraint evaluated as $\rho \to \rho_b$ then requires $C$ to satisfy
\be
 - 4\left[1-\tau_b \left(1+ \frac{3C}{2} \right)\right] \tau_b C + 3 \,\tau_b^2C^2 = {\tau_b'}^2 \,,
\ee
where we define for convenience $\tau_b :=\frac{1}{2\pi} \, \kappa^2W_b^4 T_b(\phi_b)$ and $\tau_b' :=\frac{1}{2\pi}\, \kappa^2W_b^4 T_b'(\phi_b)$.

Solving --- with root chosen so $\tau_b'\to 0$ gives $C\to 0$ --- completely determines $C$ in terms of $T_b$ and its derivative,
\be \label{Cresult}
 \tau_b \, C = - \frac{2}{9}(1-\tau_b) + \sqrt{\left( \frac{2}{9}\right)^2 (1-\tau_b)^2 + \frac{{\tau_b'}^2}{9} } \simeq \frac{{\tau_b'}^2}{2(1-\tau_b)}+\cO({\tau_b'}^4) \,,
\ee
and so shows explicitly how the $\delta$-function must depend on the metric to remain consistent with the known boundary conditions and stress-energy balance within the brane. In particular, $C$ is always nonzero whenever $T' \ne 0$
in agreement with what is found with the more elaborate but explicit UV completions of the brane source considered in \cite{Companion2}.

Notice in particular that \pref{eq:BBCnew} implies a deviation from the usual tension/defect-angle relation whenever $T'$ is nonzero.

\section{Where we stand}

The above arguments show how pressure-balance constraints dictate a brane's transverse stress-energy, $T_{mn}$, as a specific function of its tension, $T(\phi)$, and that this function generically does not vanish unless $\partial T/\partial \phi$ also does. Inferences drawn (such as those of \cite{DilatonDeltas} about the circumstances under which $\delta S_{\rm loc}/\delta \phi$ can vanish) using incomplete arguments that do not track the implications of stress-energy conservation are clearly not trustworthy. Because of \pref{4DR} this is clearly important when determining the size of the effective cosmological constant seen by an on-brane observer.

However just because a statement is not adequately supported does not make it false. Some of the conclusions of \cite{DilatonDeltas} {\em are} supported by the more detailed explorations of \cite{Companion2}, and by the determination of the 4D perspective of the low-energy 4D effective theory below the KK scale provided in \cite{Companion3}. In particular these studies do identify an important conceptual error in some of the earlier SLED papers, most notably in \cite{TNCC} and its subsequent extensions\footnote{The papers \cite{AccDistSUSY} discuss loop corrections to the background proposed in \cite{TNCC}; although the loop calculations remain valid despite this error, a more refined perspective should be adopted when considering the background about which they are computed.} \cite{AccDistSUSY}.

The important issue concerns whether or not the limit $\phi' \to 0$ as $\rho \to \rho_b$ is possible without the branes being scale invariant. This issue is important because we know from the above that the constraint ensure that if $\phi'$ vanishes in the near-brane limit then $W'$ does as well and so also does $\phi' + 2W'/W$. This suffices to ensure vanishing on-brane curvature, $\check R = 0$. But if this is also precisely scale invariant then Weinberg's no-go argument \cite{Wbgnogo} makes this less interesting by ensuring all other mass scales vanish too.

For the pure-tension branes discussed here ({\em ie} with action \pref{Sloc}) it has long been known that the brane preserves the bulk scale-invariance iff $T'(\phi) = 0$, and because $T' = 0$ implies $\phi' \to 0$ near the brane it is true that strictly vanishing $\check R$ only occurs in the scale-invariant case. Furthermore, it has been known since \cite{SLED} that the requirements of flux quantization make the bulk components of $T_{mn}$ the most dangerous for generating nonzero $\check R$. These issues are what led to the study of the interplay between tension and a brane-localized flux (BLF) term in the brane action \cite{BLFFluxQ}
\be
 S_\BLF = - \int \omega(\phi) {}^\star A \,,
\ee
where ${}^\star A$ is the 6D Hodge dual of the Maxwell field strength and the coupling function $\omega(\phi)$ is related to the amount of flux localized on the brane. Because of the metrics hidden in ${}^\star A$ the BLF interaction preserves scale invariance only if $\omega \propto e^{-\phi}$, making scale invariance appear to differ from the condition $\delta S_{\rm loc} /\delta \phi = 0$ once BLF is present, potentially opening up the possibility of having $\phi' \to 0$ (and so $\check R = 0$) without scale invariance.

This reasonable-sounding conclusion turns out to be wrong and closer examination shows that the conditions for scale invariance and $\check R = 0$ remain equivalent even with brane-localized flux. The reason for this is that although the limit $\phi' \to 0$ requires $\delta S_{\rm loc}/\delta \phi = 0$, the back-reaction of the gauge field to the presence of the BLF interaction also introduces a localized energy into the bulk Maxwell action, and it is the {\em total} localized action that must be $\phi$-independent to ensure $\check R = 0$. As proven in \cite{Companion, Companion2} (and indeed argued in \cite{DilatonDeltas}) the conditions for scale invariance and vanishing near-brane $\phi'$ agree once all sources of localized dilaton coupling are included.

Although conceptually important, it is also true that this observation does not appreciably alter the specifics of how $\check R$ depends on brane properties. This can be seen by comparing the results of \cite{Companion, Companion2, Companion3} with those of \cite{BLFFluxQ}, for the value of $\check R$ for various choices of $\phi$-dependent $S_{\rm loc}$. What it does is clarify why $\check R$ is not smaller than was found in these explicit examples.

In the end what we seek is not a precise vanishing of $\check R$ but a suppression in the low-energy cosmological consant relative to the electroweak scale, which necessarily involves breaking scale invariance. The issue is whether (and if so, by how much) $\check R$ can be suppressed by different choices for brane-bulk couplings, and if these choices can be technically natural. The first indications are \cite{Companion3} (see also \cite{BLFFluxQ}) that some suppression may be possible classically, but the re-examination of its stability to perturbations (including quantum corrections) remains incomplete.

\appendix

\section{Asymptotic forms}
\label{App:Kasner}

Bulk fields generally diverge near codimension-two (and higher) sources, and these divergences can complicate asymptotic arguments like those of \S\ref{sec:deltafail} if not treated properly. We summarize here how the near-brane solutions of \cite{6DdS} (also reproduced in \cite{Companion2}) more precisely govern the near-brane derivatives despite these divergences.

The starting point is the near-brane power-law solutions of the bulk equations near a singular source point, which can be written as a power series in the proper distance from the source, $\hat\rho := \rho - \rho_b$,
\bea \label{powerforms}
 W &=& W_0 \left( \frac{\hat\rho}{\ell} \right)^{w_b} + W_1 \left( \frac{\hat\rho}{\ell} \right)^{w_b+1} +  \cdots \nn\\
 B &=& B_0 \left( \frac{\hat\rho}{\ell} \right)^{\beta_b} + B_1 \left( \frac{\hat\rho}{\ell} \right)^{\beta_b+1} +  \cdots \\
 \hbox{and} \qquad
 e^\phi &=& e^{\phi_0} \left( \frac{\hat\rho}{\ell} \right)^{z_b} + \cdots \,, \nn
\eea
where $\ell$ is a dimensionful measure of the bulk's proper size which is by assumption much larger than the brane's size: $\ell \gg \hat\rho$. The coefficients of the series $W_i$, $B_i$ and $\phi_i$ are dictated by recursion relations arising from the bulk field equations, and these equations also impose two relations --- the Kasner conditions of \pref{Kasnerc} --- amongst the three powers $w_b$, $\beta_b$ and $z_b$ (which capture the divergent behaviour of the bulk fields near the source). In particular, the quadratic relation constrains how seriously the bulk fields can diverge by implying $w_b$, $\beta_b$ and $z_b$ must satisfy the inequalities
\be \label{limits}
 |w_b| \le \frac{1}{2} \qquad \hbox{and} \qquad
 |\beta_b |, |z_b| \le 1 \,.
\ee
The bulk field equations leave one combination of the parameters $\beta_b$, $w_b$ and $z_b$ free, and it is this free combination that is determined by the physical properties of the source, as follows.

The near-brane solution in eq.~\pref{powerforms} can be inserted into the boundary condition \pref{eq:dilatonBC} and this gives
\be \label{eq:ztprime}
 \tau^\prime_b(\phi_b)= \lim_{\rho \to \rho_b} \left[ BW^4 \phi^\prime \right] = z_b \left( \frac{ B_0 W_0^4}{\ell} \right) \left( \frac{\hat \rho}{\ell} \right)^{\beta_b+4w_b-1} = z_b \left( \frac{ B_0 W_0^4}{\ell} \right)  \,.
\ee
where the last equality uses the linear Kasner condition \pref{Kasnerc}. We similarly find from \pref{eq:BBCnew} that
\be
 1 - \tau_b \left( 1 + \frac{3C}{2}\right) = \lim_{\rho \to \rho_b} \left[ W^4 B^\prime \right] = \beta_b \left( \frac{ B_0 W_0^4}{\ell} \right) \,,
\ee
and the boundary condition from \pref{eq:WBCnew} gives
\be
 4 w_b \left( \frac{ B_0 W_0^4}{\ell} \right) = -2 \tau_b \, C \,.
\ee
Notice that this is always consistent with the constraint equation \pref{eq:const} because of the quadratic Kasner condition \pref{Kasnerc}.

As in the main text we see that a tacit assumption that $C = 0$ (as made so seductive in the $\delta$-function approach) immediately implies $w_b = 0$, from which the Kasner conditions then give $\beta_b = 1$ and $z_b = 0$; in manifest constradiction with \pref{eq:ztprime}. It was precisely to nail down this problem that the UV completions in \cite{Companion2} were constructed, allowing these asymptotic arguments to be tested in detail numerically.

\section*{Acknowledgements}

We acknowledge Florian Niedermann and Robert Schneider for collaborations at early stages of this work as well as many discussions about papers \cite{Companion, Companion2} and \cite{Companion3} while they were in preparation. This research was supported in part by funds from the Natural Sciences and Engineering Research Council (NSERC) of Canada, and by a postdoctoral fellowship from the National Science Foundation of Belgium (FWO), by the Belgian Federal Science Policy Office through the Inter-University Attraction Pole P7/37, the European Science Foundation through the Holograv Network, and the COST Action MP1210 `The String Theory Universe'. Research at the Perimeter Institute is supported in part by the Government of Canada through Industry Canada, and by the Province of Ontario through the Ministry of Research and Information (MRI).

\end{document}